\documentclass[aps,12pt,showpacs,showkeys]{revtex4}

\usepackage[english]{babel}
\usepackage{amsmath}
\usepackage{amssymb}
\usepackage{amsbsy}
\usepackage{amstext}
\usepackage{graphicx}
\usepackage{pst-node}
\usepackage{verbatim}

\newcommand{\be}{\begin{eqnarray}}
\newcommand{\ee}{\end{eqnarray}}
\newcommand{\bdm}{\begin{displaymath}}
\newcommand{\edm}{\end{displaymath}}
\newcommand{\ds}{\displaystyle}
\newcommand{\ba}{\begin{array}}
\newcommand{\ea}{\end{array}}
\newcommand{\pa}[1]{\left(#1\right)}
\newcommand{\paq}[1]{\left[#1\right]}
\newcommand{\pag}[1]{\left\{#1\right\}}
\newcommand{\dpa}{\partial}
\newcommand{\K}{{\bf k}}
\newcommand{\Q}{{\bf q}}

\newcommand{\X}{{\bf x}}

\begin{document}

\title{Tail terms in gravitational radiation reaction via effective field 
theory}

\author{Stefano Foffa$^{\rm 1}$ and Riccardo Sturani$^{\rm 2,3}$}

\affiliation{$(1)$ D\'epartement de Physique Th\'eorique and Center for Astroparticle Physics, Universit\'e de 
             Gen\`eve, CH-1211 Geneva, Switzerland\\
             $(2)$ Dipartimento di Scienze di Base e Fondamenti, 
             Universit\`a di Urbino, I-61029 Urbino, Italy\\
             $(3)$ INFN, Sezione di Firenze, I-50019 Sesto Fiorentino, Italy}

\email{stefano.foffa@unige.ch, riccardo.sturani@uniurb.it}

\begin{abstract}
Gravitational radiation reaction affects the dynamics of gravitationally bound
binary systems. Here we focus on the leading ``tail" term which modifies binary
dynamics at fourth post-Newtonian order, as first computed by Blanchet and 
Damour.
We re-produce this result using effective field theory techniques in the 
framework of the Lagrangian formalism suitably extended to include dissipation
effects. We recover the known logarithmic tail term, consistently with the 
recent interpretation of the logarithmic tail term in the mass parameter as a 
renormalization group effect of the Bondi mass of the system.
\end{abstract}



\maketitle

\section{Introduction}

The forces induced on a isolated system by reaction to the emission of
gravitational waves and their impact on the motion of gravitationally bound 
binary systems have been studied in great accuracy since the first derivation of
the radiation reaction force in General Relativity by Burke and Thorne 
\cite{BurkeThorne}.
Their phenomenological impact is linked to the forthcoming
observation runs of the Laser Interferometer Gravitational Observer (LIGO)
and Virgo, see \cite{:2012dr} for the result of the latest compact binary
coalescence search, and have already been observed to be at play in 
binary pulsar systems \cite{Hulse:1974eb,Taylor:1982zz}.

The motion of coalescing binaries is imprinted in the shape of the emitted
gravitational waves and the output of gravitational detectors is particularly
sensitive to the time varying phase of the radiated wave, which has to be
determined with high accuracy in order to ensure high efficiency of the 
detection algorithms and faithful source parameter reconstruction.\\
The standard approach to describe the motion of coalescing binaries lies
within the post-Newtonian (PN) approximation to General Relativity, 
describing the binary system dynamics as a perturbative series in terms of the 
relative velocity of the binary constituents, see e.g. \cite{Blanchet_living} 
for a review.

The leading effect of radiation reaction modifies the binary dynamics
giving rise to a term non-invariant under time reversal which affects the 
dynamics of the system at 2.5 PN order: this is the lowest order at which 
linear effects of the gravitational radiation enter.\\
In \cite{Blanchet:1987wq,Blanchet:1993ng} the leading 
non-linear radiation reaction effect has been derived, 
see also \cite{Blanchet:1993ec}, and  it is shown to 
modify the binary dynamics at 4PN order (i.e. at 1.5PN order relatively to the 
leading effect): it belongs to the species of terms dubbed \emph{hereditary}, 
as it depends on the entire history of the source.
In particular it originates from radiation emitted and then scattered back into 
the system by the background curvature generated by the total mass $M$ of the 
binary system, hence the name of \emph{tail} term. Such non-linear 4PN tail 
term is in good agreement with computations performed within the 
framework of the gravitational self-force analysis of circular orbits in 
Schwarzschild background, as found in \cite{Blanchet:2010cx,Blanchet:2010zd} 
and it includes both a term non-invariant and a term \emph{invariant} under 
time reversal; the latter can be incorporated in the conservative dynamics of 
the binary system.

Here we present the re-computation of the 4PN tail contribution to the 
dynamics of inspiraling binaries via the use of the Effective
Field Theory (EFT) methods for gravity introduced in \cite{Goldberger:2004jt}.
EFT methods turn out to be useful in problems admitting a clear scale 
separation: in the binary system case we have the size of the compact
objects $r_s$, the orbital separation $r$ at which the system consists of
point particles interacting via instantaneous potential, and the gravitational 
wave-length $\lambda$ 
(with hierarchy $r_s < r \sim r_s/v^2 < \lambda \sim r/v$, being $v$ the 
relative velocity between the two bodies) at which the 
binary system can be described as a particle of negligible size endowed with 
multipoles.
 
Using this approach several different groups have re-produced results in the
PN analysis which have been previously computed in the standard approach
both in the conservative \cite{EFToldcons,Foffa:2011ub} and in the dissipative 
\cite{Goldberger:2009qd,EFTolddis} sector, and EFT methods have also been 
applied within the extreme mass ratio limit approach to binary coalescence 
\cite{EFTextmass}.
Moreover new results on the PN analysis have been made available by the use of
the EFT method in both sectors 
\cite{EFTnewcons,EFTnewdis,Porto:2012as,Goldberger:2012kf,Foffa:2012rn}, 

The leading orders in the radiation reaction effects has been re-produced via 
effective field theory in \cite{Galley:2009px, Galley:2012qs}, extending
the Lagrangian formalism to include time-asymmetric systems, see 
\cite{Galley:2012hx} for a rigorous extension of Hamilton's principle to 
generally dissipative systems.

\begin{center}
  \begin{figure}[t]
    \includegraphics[width=.45\linewidth]{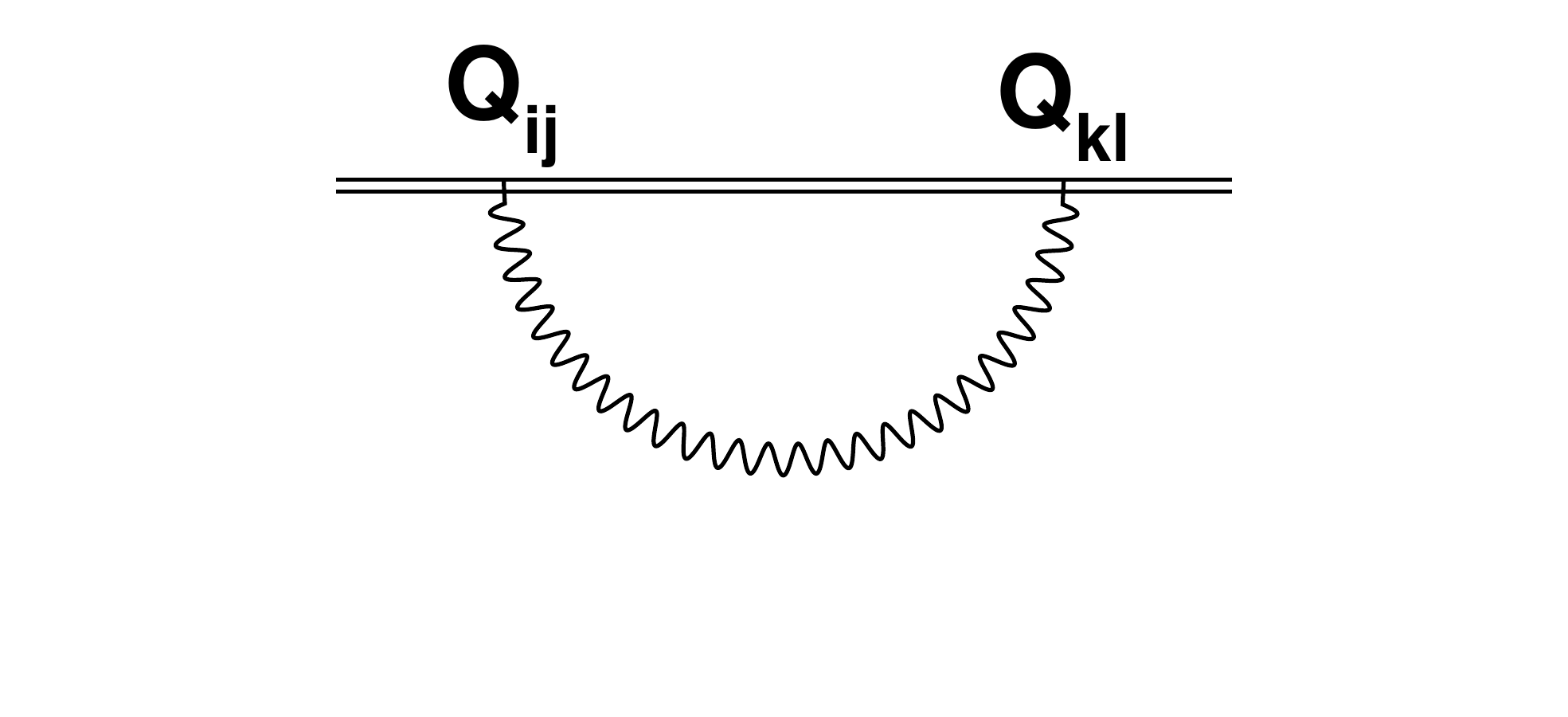}
    \includegraphics[width=.45\linewidth]{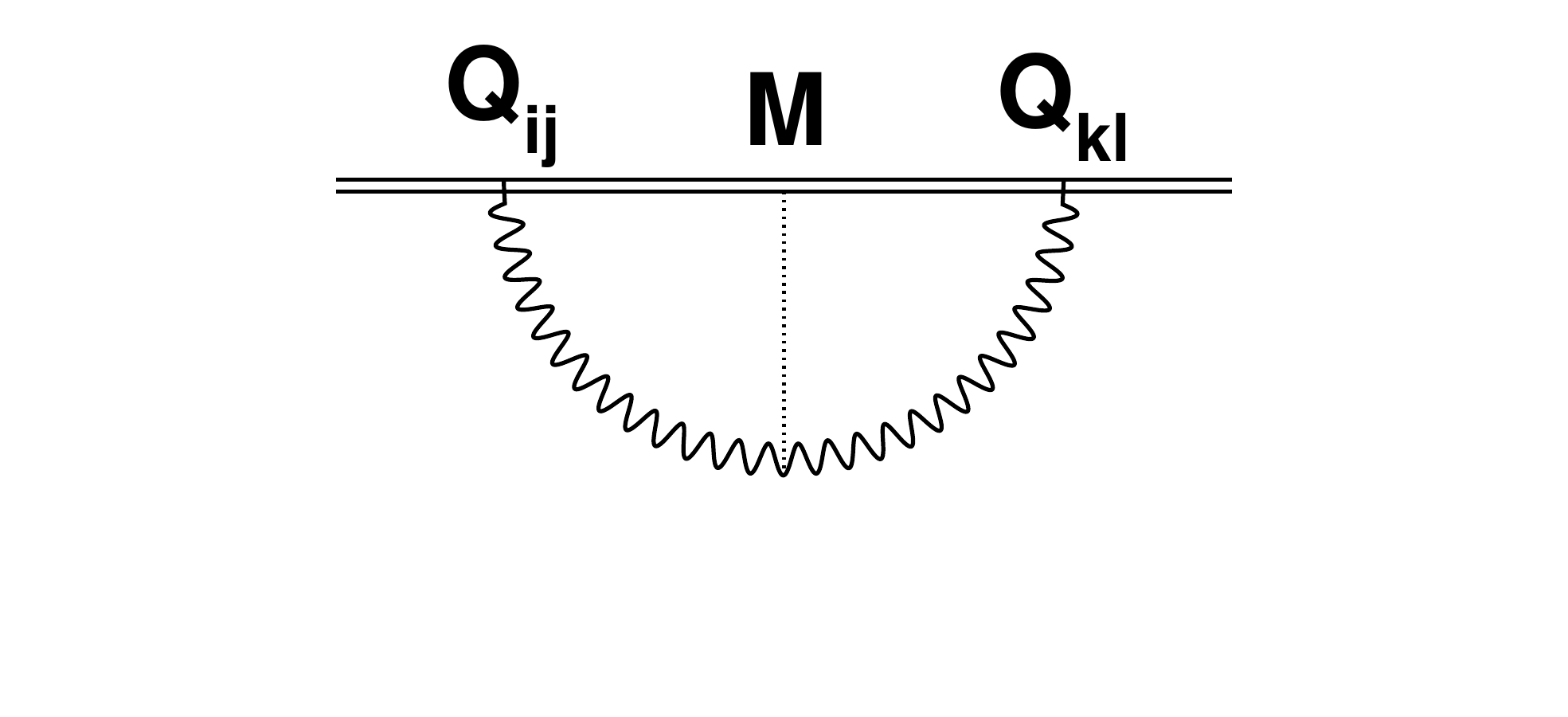}
    \caption{Diagram describing the gravitational radiation reaction force 
      at leading order (left) and leading non-linear order (right).
      The thick line represents the massive binary system, the curly line the
      the gravitons emitted and absorbed by the system, the dashed line the
      potential graviton responsible for the Newtonian potential.}
    \label{fig:radReac}
  \end{figure}
\end{center}

\section{Radiation reaction logarithms from tail term}
\label{se:QQ}

In the following we use $c=1$ units and the mostly plus signature convention.
Contraction of space indices are taken with the Kronecker delta.
We work in generic $d$ space dimensions as we adopt dimensional regularization 
to handle divergences. It will be convenient to use the $d+1$-dimensional
Planck mass $\Lambda\equiv (32\pi G)^{-1/2}$, being $G$ the $d+1$-dimensional 
gravitational constant.

Following the EFT framework for non relativistic General Relativity, after 
integrating out the ``potential'' gravitons, one is left with an effective 
action at the orbital scale $r$, describing radiation gravitons coupled to 
the multipole moments of the compact binary system.
In this limit, by adopting the decomposition of the metric suggested in 
\cite{Kol}
\be
\label{eq:metric}
g_{\mu\nu}=e^{2\phi/\Lambda}\pa{
\ba{cc}
-1 & A_j/\Lambda\\
A_i/\Lambda & \quad e^{-c_d\phi/\Lambda}\pa{\delta_{ij}+\sigma_{ij}/\Lambda}-A_iA_j/\Lambda^2
\ea}\,,
\ee
with $c_d\equiv 2\frac{(d-1)}{(d-2)}$, the dynamics is described by an 
effective word-line Lagrangian coupled to gravity, whose relevant terms for the 
present work are
\be
\label{eq:smult}
S_{mult}\supset -\int d\tau \pa{M+\frac{M\phi}{\Lambda}-\frac 12 Q_{ij}R^0_{\ i0j}}
\,,
\ee
where $R^0_{\ i0j}$ denotes the appropriate component of the Riemann tensor,
and $M$ and $Q_{ij}$ are, respectively, the mass monopole and quadrupole moments
of the system;
dependence on the time-like coordinate parametrizing the word-line is 
understood in all terms in $S_{mult}$. The bulk dynamics of the gravitational
fields $\phi$, $A$ and $\sigma$ is given by the standard Einstein-Hilbert 
action plus gauge fixing, whose terms relevant for the present calculation are 
reported in the appendix.

In order to perform the computation of the diagrams of fig.~\ref{fig:radReac}, boundary conditions 
asymmetric in time have to be imposed, as no incoming radiation at past null 
infinity is required. Technically this is implemented 
by adopting a generalization of the Hamilton's variational principle
similar to the closed-time-path, or in-in formalism (first
proposed in \cite{Schwinger:1960qe}, see \cite{deWitt} for a review) 
as described in \cite{Galley:2012hx}, which requires a doubling of the fields 
variable. For instance, for a free scalar field $\Psi$, the generating 
functional $W$ for connected correlation functions in the in-in formalism has 
the path integral representation
\renewcommand{\arraystretch}{1.5}
\be
\label{eq:doubleP}
\ba{rcl}
\ds e^{i\mathcal{S}_{eff}[J_1,J_2]}&=&\ds\int \mathcal{D}\Psi_1\mathcal{D}\Psi_2 
\exp\pag{\int d^{d+1}x
\,\paq{-\frac i2(\dpa\Psi_1)^2+\frac i2(\dpa\Psi_2)^2-iJ_1\Psi_2+iJ_2\Psi_2}}\,.
\ea
\ee
\renewcommand{\arraystretch}{1.}
In this toy example the path integral can be performed exactly, and using the 
Keldysh representation \cite{Keldysh:1964ud} defined by
$\Psi_-\equiv\Psi_1-\Psi_2$, $\Psi_+\equiv (\Psi_1+\Psi_2)/2$,
one can write
\be
\mathcal{S}_{eff}[J_+,J_-]=\frac i2\int d^4x\,d^4y J_B(x)G^{BC}(x-y)J_C(y)\,,
\ee
where the $B,C$ indices take values $\{+,-\}$ and
\be
\label{eq:prop}
G^{BC}(t,\X)=
\pa{\ba{cc}
0 & -iG_A(t,\X)\\
-iG_R(t,\X) & \frac 12 G_H(t,\X)
\ea}\,,
\ee
where $G^{++}=0$ and $G_{A,R,H}$ are the usual advanced, retarded propagators
and Hadamard function respectively, see sec.~\ref{se:apGreen} for more detailed 
formulae.
In our case, the expression of the quadrupole in terms of the binary 
constituents world-lines $\X_a$, i.e.
\be 
Q_{ij}\equiv\sum_{a=1}^2 m_a\pa{\X_{ai}\X_{aj}-\frac{\delta_{ij}}d\X_{ak}\X_{ak}}\,,
\ee
is doubled to
\renewcommand{\arraystretch}{.6}
\be
\ba{rcl}
\ds Q_{-ij}&=&\ds\sum_{a=1}^2m_a\pa{x_{-ai}x_{+aj}+x_{+ai}x_{-aj}}
-\frac 2d\delta_{ij}x_{+ak}x_{-ak}\\
\ds Q_{+ij}&=&\ds\sum_{a=1}^2m_ax_{+ai}x_{+aj}-\frac 1d\delta_{ij}x_{+a}^2+O(x_-^2)\,.
\ea
\ee
\renewcommand{\arraystretch}{1.}
The word-line equations of motion that properly include radiation reaction 
effects are given by
\be
\label{eq:eqmoto}
\left.0=\frac{\delta S_{eff}[\X_{1\pm},\X_{2\pm}]}{\delta \X_{a-}}\right|_{\substack{\X_{a-}=0\\ \X_{a+}=\X_a}}\,.
\ee

At lowest order, by integrating out the radiation graviton, i.e. by computing 
the diagram in the left of fig.~\ref{fig:radReac}, the Burke-Thorne potential
\cite{BurkeThorne} is obtained from the action
\be
\label{eq:BT}
S_{eff}^{(Q^2)}=-\frac{G_N}5\int dt\,Q_{-ij}(t)Q^{(5)}_{+ij}(t)\,,
\ee
where $A^{(n)}(t)\equiv d^nA(t)/dt^n$, and $G_N$ the standard Newton's constant, 
which has been derived in the EFT framework in \cite{Galley:2009px}. 
Corrections to the leading effect appears at relative 1PN
order due to the inclusion of higher multipoles and the 1PN modified dynamics of
the quadrupole \cite{355730,Galley:2012qs}.
The genuinely non-linear effect appear at relative 1.5PN order and it is due
to the rightmost diagram in fig.~\ref{fig:radReac}.

In order to compute the $S_{eff}$ we expand the metric as in 
eq.~(\ref{eq:metric}) and integrate out the fluctuations $\phi,A,\sigma$ 
according to the in-in prescription, getting to an effective action
\be
\label{eq:Wick}
iS_{eff}[M,Q_\pm]=\int D\phi_\pm D\sigma_\pm DA_\pm e^{iS(\phi_\pm,\sigma_\pm,A_\pm,M,Q_\pm)}\,,
\ee
for the multipole moments alone (we have denoted by
$S(\phi_\pm,\sigma_\pm,A_\pm,M,Q_\pm)$ the action including both the standard 
Einstein-Hilbert action (plus gauge-fixing) and the $S_{mult}$ from 
eq.~(\ref{eq:smult})).
The diagram on the right of fig.~\ref{fig:radReac}, see sec.~\ref{ss:coord} for 
computation details, gives the following logarithmic contribution to the 
effective action (by virtue of eq.~(\ref{eq:eqmoto}) only terms linear in $Q_-$
are kept)
\be
\label{eq:radReacT}
S_{eff}^{(MQ^2)}=-\frac 45G_N^2M\int dt\, Q_{-ij}(t)\int_{-\infty}^t dt'Q_{+ij}^{(6)}\,
\frac 1{(t-t')}
\ee
which exhibits a short distance singularity for the gravitational wave being
emitted and absorbed at the same space-time point (with Green functions used
in their $d=3$ expression).
Actually the complete result of the tail diagram in fig.~\ref{fig:radReac} in 
momentum space, dimensionally regularized, reads
\renewcommand{\arraystretch}{1.4}
\be
\label{eq:radReacRes}
\ba{rcl}
S_{eff}^{(MQ^2)}&=&\ds -\frac 15G_N^2M\int_{-\infty}^\infty\frac{dk_0}{2\pi}\,k_0^6
\ds\pa{\frac 1\epsilon-\frac{41}{30}+i\pi-\log\pi+\gamma+
\log(k_0^2/\mu^2)}\times\\
&&\ds\paq{Q_{ij-}(k_0)Q_{ij+}(-k_0)+Q_{ij-}(-k_0)Q_{ij+}(k_0)}\,,
\ea
\ee
\renewcommand{\arraystretch}{1.}
where we the $3+1$-dimensional gravitational constant $G_N$ is related to the 
one in generic space dimension $d$ by $G=1/(32\pi\Lambda^2)=G_N\mu^{-\epsilon}$.
By performing the computation in $d=3+\epsilon$ the logarithmic divergence has 
been regularized and a spurious dependence on the arbitrary subtraction 
scale $\mu$ has been introduced
\footnote{The logarithmic term is non-analytic in $k_0$-space and non-local 
(but causal) in direct space: after integrating out a mass-less propagating 
degree of freedom the effective action does not have to be local, and indeed it
is not.}.
A local counter term $M_{ct}$ defined by
\be
M_{ct}=-\frac{2G_N^2}5M\pa{\frac 1\epsilon +\gamma -\log\pi}Q_{-ij}Q_{+ij}^{(6)} 
\ee
can be straightforwardly added to the world-line effective
action to get rid of the divergence appearing as $\epsilon\to 0$. 
According to the standard renormalization procedure, one can define a 
renormalized mass $M^{(R)}(t,\mu)$  for the monopole term in the action 
(\ref{eq:smult}), depending on time (or frequency) and on the arbitrary scale 
$\mu$ in such a way that physical quantities (like the energy or the radiation 
reaction force) will be $\mu$-independent.

Note that at the order required in the diagram in fig.\ref{fig:radReac}, 
$M^{(R)}(t,\mu)$ can be safely treated as a constant $M$ on both its arguments 
$t$ and $\mu$.  Also the renormalization of the quadrupole moment, which occurs
at 3PN order with respect to its leading value, see \cite{Goldberger:2009qd}, 
can be neglected here.

From eq.~(\ref{eq:radReacRes}), representing the contribution to the radiation
reaction force of the tail diagram, the multipole effective action relevant from
the tail process can be derived to be:
\renewcommand{\arraystretch}{1.4}
\be
\label{eq:radReacRen}
\ba{rcl}
\ds S_{eff}&=&\ds\int\frac{dk_0}{2\pi}\left\{M^{(R)}(k_0,\mu)
-i\frac{G_N}{5}k_0^5Q_{-ij}(-k_0)Q_{+ij}(k_0)\right.\\
&&\ds \!\!\!\!\!\! \left.-\frac{G_N^2}5M\,k_0^6
\paq{\log(k_0^2/\mu^2)-\frac{41}{30}+i\pi}
\pa{Q_{ij-}(k_0)Q_{ij+}(-k_0)+Q_{ij-}(-k_0)Q_{ij+}(k_0)}\right\}\,,
\ea
\ee
\renewcommand{\arraystretch}{1.}
where the renormalized monopole term appears: we will now determine its explicit
$\mu$ dependence, recovering the result found in \cite{Goldberger:2012kf}, by 
requiring that the physical energy does not depend on $\mu$.

\begin{figure}
  \begin{center}
    \includegraphics[width=.32\linewidth]{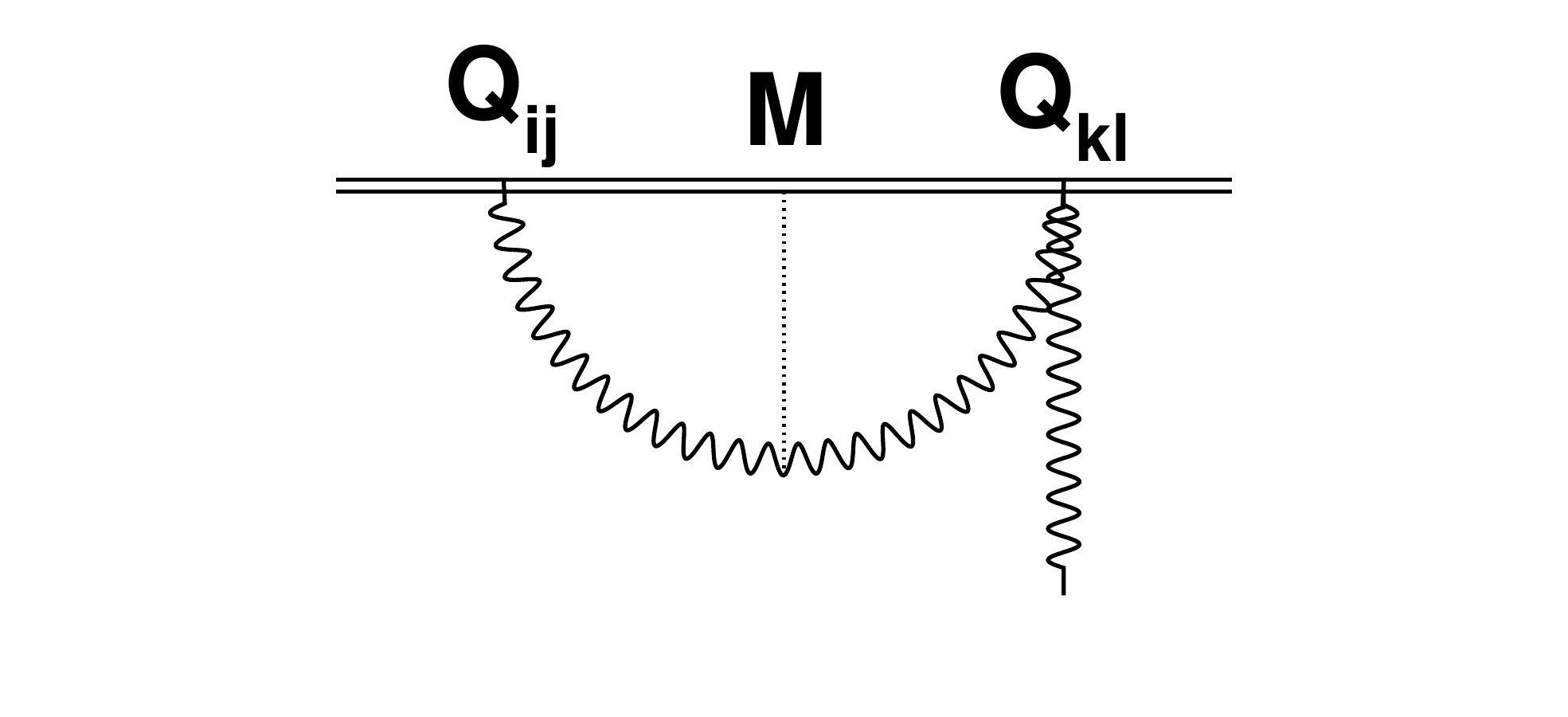}
    \includegraphics[width=.32\linewidth]{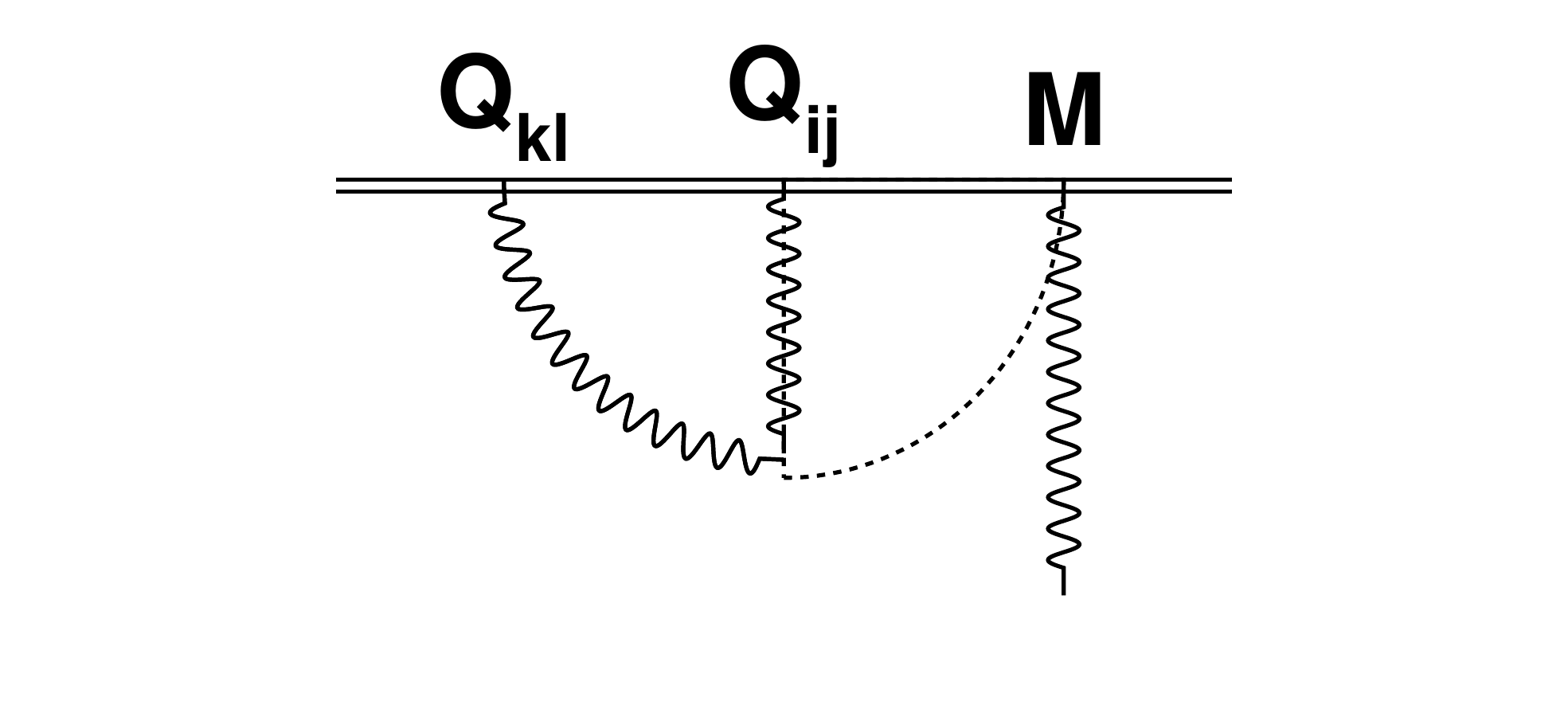}
    \includegraphics[width=.32\linewidth]{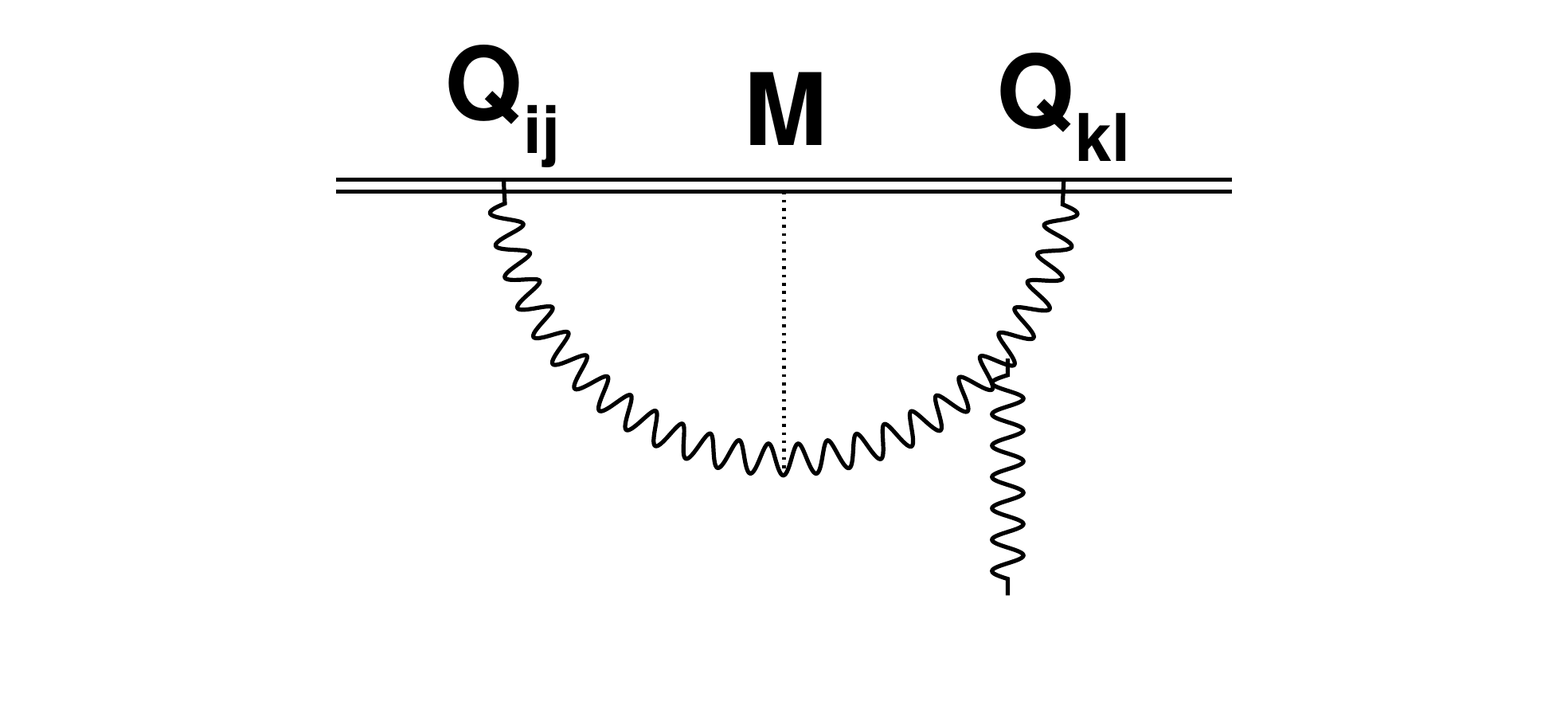}\\
    \includegraphics[width=.32\linewidth]{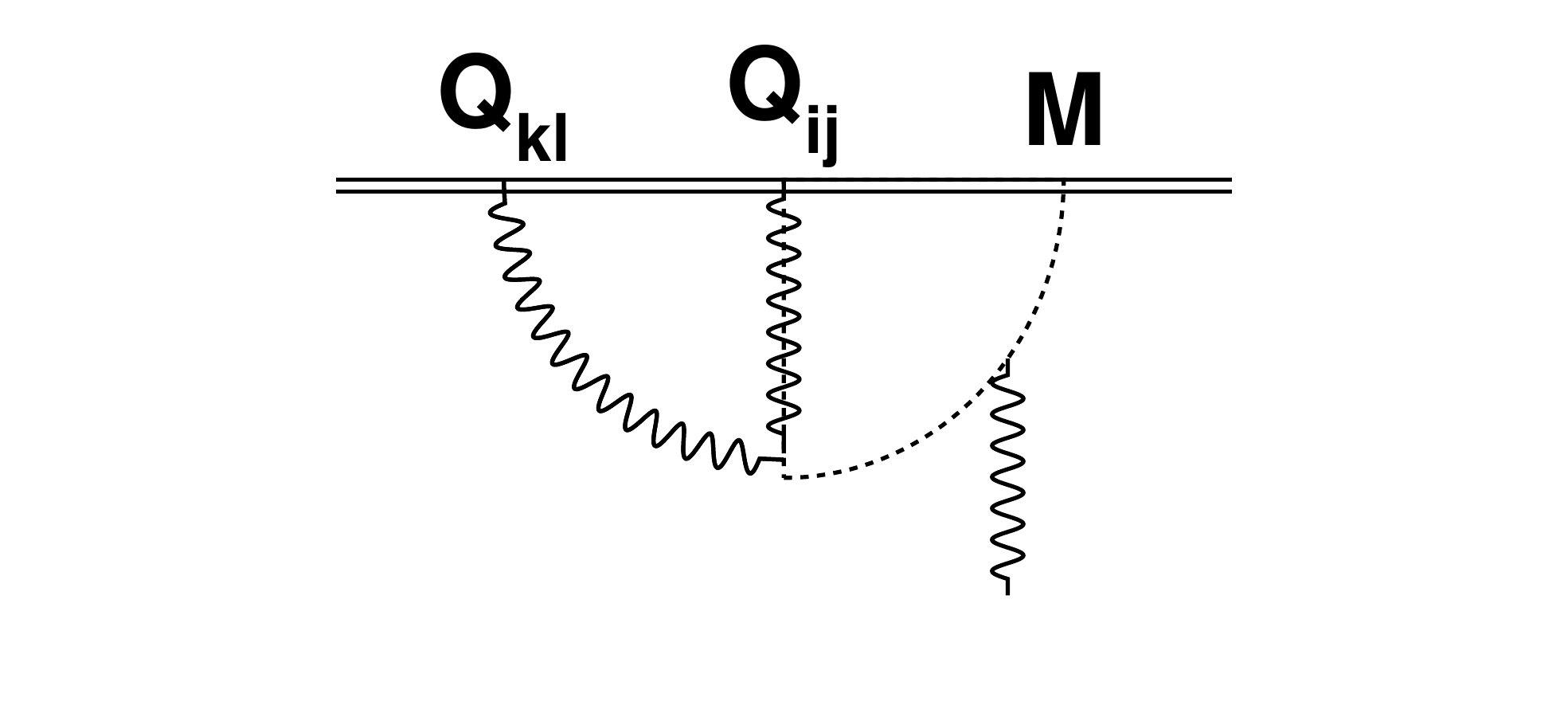}
    \includegraphics[width=.32\linewidth]{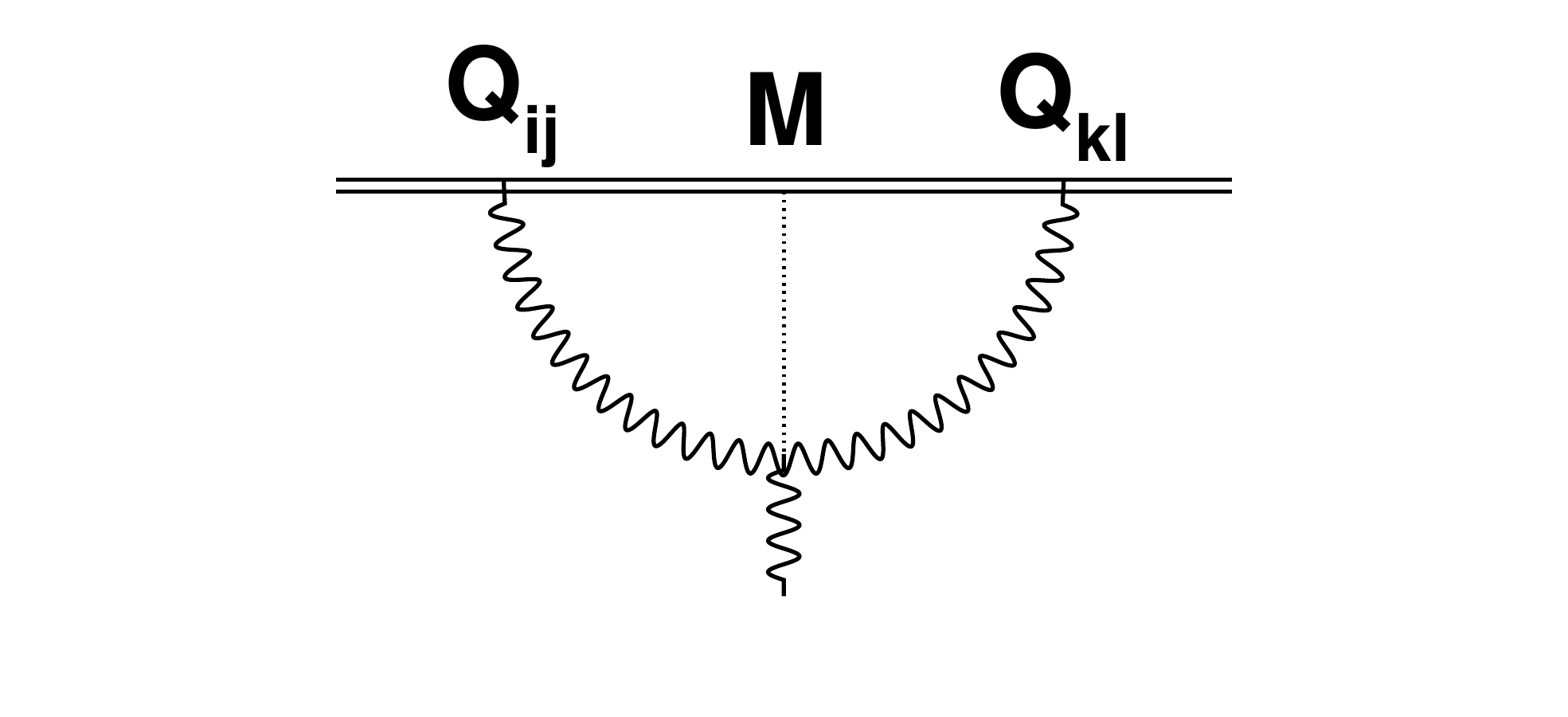}
  \end{center}
\caption{Series of diagram studied in \cite{Goldberger:2012kf} showing that the
mass monopole undergoes a non-trivial renormalization group flow.}
\label{fig:massRen}.
\end{figure}

We start by deriving the contribution that the effective action 
(\ref{eq:radReacRen}) makes to the equations of motion of the binary 
constituents, limiting to the logarithmic term
\be
\label{eq:resreg}
\left.\delta\ddot x_{ai}(t)\right|_{log}=
-\frac 85 x_{aj}(t)G_N^2 M\int^t_{-\infty}dt'\,Q_{ij}^{(7)}(t')\log\paq{(t-t')\mu}\,,
\ee
which agrees with the result obtained in 
\cite{Blanchet:1993ng}. Note that the normalization of the time (i.e. the value
of $\mu$) in the logarithm is arbitrary: changing the time normalization shifts
the action by a quantity proportional to $x_{aj}(t)G_N^2 M Q_{ij}^{(6)}(t)$,
see next section for a discussion of such analytic, local term.

Following \cite{Blanchet:2010zd}, we can separate the logarithm argument into a 
$t$-dependent and a $t$-independent part via the trivial identity
\be
\log\paq{(t-t')\mu}=\log\paq{(t-t')/\lambda}+\log\pa{\lambda\mu}\,,
\ee
for any $\lambda$.
The logarithmic term not-involving time gives a \emph{conservative} 
contribution to the force in eq.~(\ref{eq:resreg}) which gives a logarithmic 
shift $\delta M$ to the mass of the binary system. The mass-shift $\delta M$
can be determined by observing that 
\be
\frac{d(\delta M^{(R)})}{dt}=-\sum_a m_a\delta\ddot x_{ai}\dot x_{ai}
\ee
and thus the tail contribution tail contribution to the conservative part of 
the energy $E$ is
\cite{Blanchet:2010zd} 
\be
\label{eq:dE}
E=M^{(R)}+\sum_a m_a\delta\ddot x_{ai}x_{ai}=M^{(R)}+\frac{2G_N^2M}5\pa{2Q_{ij}^{(5)}Q_{ij}^{(1)}-2Q_{ij}^{(4)}Q_{ij}^{(2)}+Q_{ij}^{(3)}Q_{ij}^{(3)}}\log(\mu\lambda)\,,
\ee
where the renormalized monopole term has also been included, as it gives a
contribution of the order $G_NMQ^2$: actually by imposing
that the physical energy $E$ does not depend on $\mu$ one can find the 
renormalization group flow equation
\be
\mu\frac{d}{d\mu}M(t,\mu)=-\frac{2G_N^2M}5\pa{2Q_{ij}^{(5)}Q_{ij}^{(1)}
-2Q_{ij}^{(4)}Q_{ij}^{(2)}+Q_{ij}^{(3)}Q_{ij}^{(3)}}\,,
\ee
which agrees with the result found in \cite{Goldberger:2012kf}, where the 
monopole mass $M$ (identified with the Bondi mass of the binary system)
is shown to undergo a non-trivial renormalization group flow by analyzing the 
diagrams in fig.~\ref{fig:massRen}.

\section{Finite quantity from tail terms}

What about the finite part? The divergence encountered in the previous section
comes from the lack of UV-completeness of the effective model when treating 
the coalescing binary as a fundamental system endowed with multipoles moments: 
the exact numerical result is sensitive to the short distance physics and the 
EFT in terms of source multipoles does not know about it.
Such numerical quantity can be fixed by performing the radiation reaction 
computation in the full theory of gravity coupled to individual (point-like) 
binary constituents.

\begin{center}
  \begin{figure}
    \includegraphics[width=.6\linewidth]{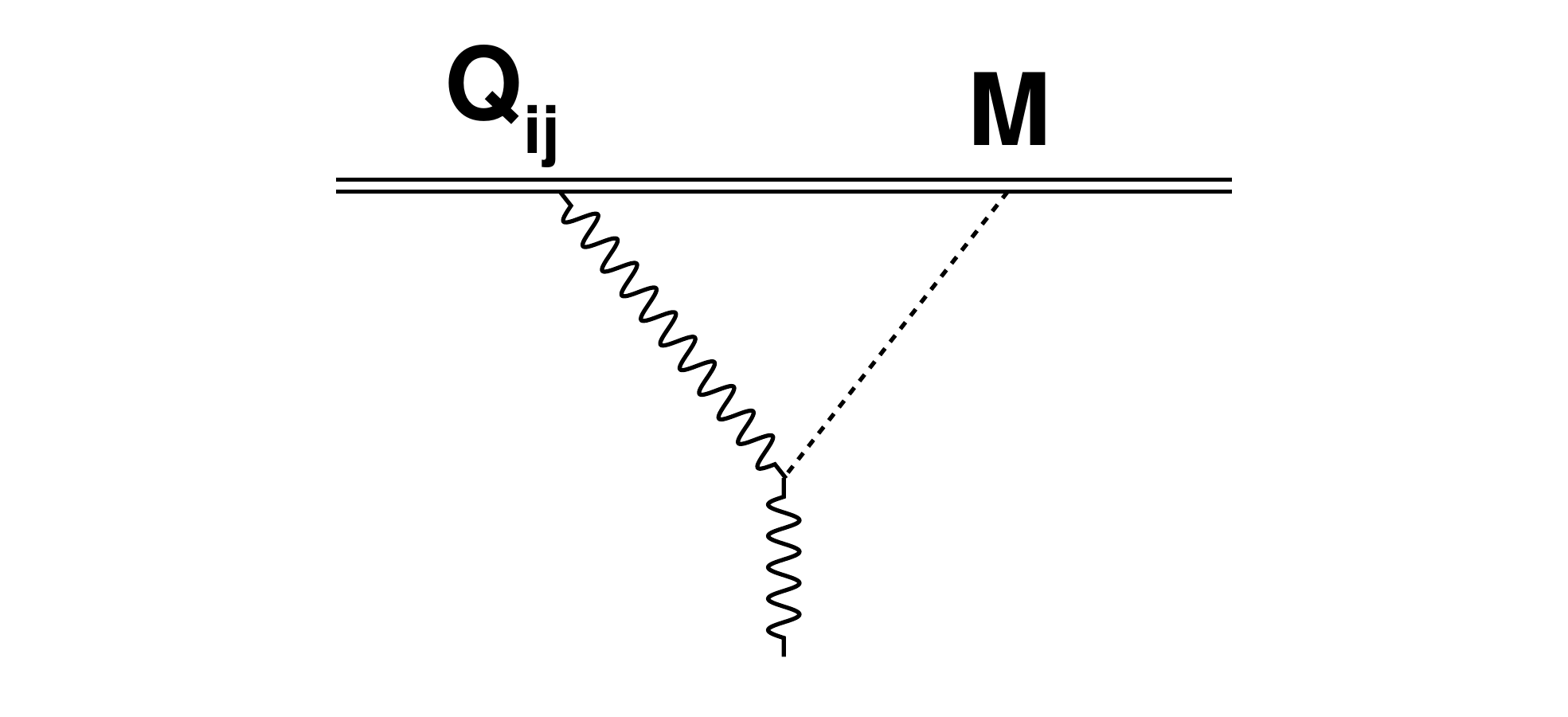}
    \caption{Diagram describing the gravitational radiation emitted by a 
      quadrupole source and scattered off the Newtonian potential before
      escaping to infinite.}
    \label{fig:htail}
  \end{figure}
\end{center}

Within the traditional approach, the finite analytic term entering the radiation
reaction force was actually computed in \cite{Blanchet:1993ng}, by relating the
radiation reaction potential to the ``anti-symmetric'' (i.e. non-time invariant)
wave perturbation of the time-time component of the metric generated by the 
quadrupole, which was in turn fixed to the $ij$ component.
The gravitational wave in the Trasverse-Traceless gauge, including the 
tail effect, has been computed in \cite{Blanchet:1993ng,Blanchet:1993ec} to be:
\be
\label{eq:htail}
h^{(TT)}_{ij}=-\Lambda_{ij,kl}\frac{2G_NM}r\int\frac{dk_0}{2\pi} 
e^{ik_0(t-r)}k_0^2Q^{(tail)}_{kl}(k_0)\,,
\ee
with 
\be
\label{eq:qTail}
Q^{(tail)}_{kl}\equiv Q_{kl}\pag{1+G_NMk_0
\paq{-2i\pa{\frac 1\epsilon+\log(k_0/\mu)+
\frac{\gamma}2-\frac{11}{12}}+\pi sgn(k_0)}},
\ee
showing a long-scale (IR) divergence due to the gravitational wave 
emitted by the quadrupole source and scattered off the by the long-ranged 
Newtonian potential. The IR singularity in the phase of the emitted wave is
un-physical as it can be absorbed in a re-definition of time in 
eq.~(\ref{eq:htail}) by exponentiation the imaginary term in 
eq.~(\ref{eq:qTail}). Moreover any experiment, like LIGO and Virgo for 
instance, can only probe phase \emph{differences} (e.g. the gravitational wave 
phase difference between the instants when the wave enters and exits the 
experiment sensitive band) and the un-physical dependencies on the regulator 
$\epsilon$ and on the subtraction scale $\mu$ drops out of any observable.
Such result has been re-derived within EFT techniques in \cite{Porto:2012as}
by computing the diagram in fig.~\ref{fig:htail}.

Actually the diagram computed in the previous section is related to
the one in fig.~\ref{fig:htail}.
In order to recover the right diagram in fig.~\ref{fig:radReac} from 
fig.~\ref{fig:htail}, the gravitational wave emitted  has to be absorbed via
another quadrupole insertion. In this process, the IR singularity of 
fig.~\ref{fig:htail} is turned into the UV one of fig.~\ref{fig:radReac}, which occurs
when the time difference between emission and absorption goes to zero.

In \cite{Blanchet:1993ng} the radiation reaction potential corrected for the 
tail term was computed by observing that the tail effect amounts at shifting 
the quadrupole as per eq.~(\ref{eq:qTail}), and the radiation reaction 
potential can then be inferred by evaluating the radiation-reaction Burke Thorne
term in eq.~(\ref{eq:BT}) on the ``shifted'' quadrupole moment given by 
eq.~(\ref{eq:qTail}).
Such procedure gives the correct logarithmic term for the radiation reaction
force, enabling to fix the finite term analytic in $k_0$.

The finite piece in the tail term radiation reaction force
is responsible for a conservative force at 4PN (as the leading radiation 
reaction acts at 2.5PN and the tail term is a 1.5PN correction to it), so it 
must be added to the conservative dynamics coming from the calculation
of the effective action not involving gravitational radiation, 
see \cite{Foffa:2012rn,Jaranowski:2012eb} for partial results at 4PN. 
In particular, the conservative part in the radiation reaction force affects 
the (coordinate transformation invariant) energy of circular orbits $E(x,\nu)$, 
being $x$ the PN expansion parameter $x\equiv (G M\omega)^{2/3}$ and $\omega$ 
the angular frequency of circular orbits. Such energy function depends also on
the symmetric mass ratio parameter $\nu\equiv m_1 m_2/M^2$ and linearly on the 
total mass $M$, the tail effect gives a contribution proportional to $\nu^2$ to
the 4PN energy of circular orbit $E_{4PN}(x,\nu)$ (the $O(\nu)$ contribution
is known from the Schwarzschild limit). 

The complete $\nu^2$ term of $E_{4PN}(x,\nu)$ have been computed in 
\cite{LeTiec:2011ab} within the context of the extreme mass ratio inspiral 
approximation (where $\nu$ is the expansion parameter and the metric expanded 
around the curved background created by the more massive object forming the 
binary system), with the result
\be
\left.E_{4PN}(x,\nu)\right|_{\nu^2}=-\frac 12\nu^2 M x^5
\pa{e_1+\frac{448}{15}\log(x)}\,,
\ee
with $e_1\simeq 153.8803$ \cite{LeTiec:2011ab}.
The logarithmic term matches the term derived from the PN approximation, with
traditional method as done in \cite{Blanchet:2010zd}, and EFT methods: both via
the computation of the mass renormalization as done in \cite{Goldberger:2012kf}
or via the computation of the radiation reaction force as done here.
Work is under-way to derive the full $E_{4PN}(x,\nu)$ in a PN context.

\section{Conclusion}
The conservative dynamics of gravitationally bound binary systems is completely 
known in literature up to the third post-Newtonian order, result being derived
by both traditional methods and within the context of effective field 
theory methods. At fourth post-Newtonian order the conservative dynamics 
receives contribution from a process involving the emission and absorption of 
radiation (a so-called radiation reaction process), giving rise to logarithmic 
and analytic terms in the post-Newtonian expansion parameter. 
We have computed here the logarithmic part of the radiation reaction potential,
affecting both the dissipative and the conservative dynamics, 
using effective field theory methods, and comparing the result with related
ones obtained with different methods within the post-Newtonian framework.
The fourth post-Newtonian order contribution to the conservative dynamics, at 
specific order in the symmetric mass ratio, which includes the radiation 
reaction tail term, has been computed
in literature within the extreme mass ratio inspiral approximation and its
logarithmic part agrees with what has been computed within the framework of the 
post-Newtonian approximation to General Relativity.
Work is under-way within the post-Newtonian approximation to recover the full
fourth order energy function, including all the terms analytic in the 
post-Newtonian expansion parameter.

\section*{Acknowledgments}
The authors wish to thank G. Cella and R. Porto for useful discussions.
SF wishes to thank the Dipartimento di Scienze di Base e Fondamenti of the
University of Urbino for kind hospitality during the realization of part of 
this work, RS wishes to thank the D\'epartment de Physique Th\'eorique of the 
University of Geneva and the Instituto de F\'\i sica Te\'orica of the UNESP of 
Sao Paulo for kind hospitality and support during the realization of part of 
this work.
The work of SF is supported by the Fonds National Suisse, the work of RS is 
supported by the EGO Consortium through the VESF fellowship EGO-DIR-41-2010.

\appendix
\section{Useful formulae}

\subsection{Green functions}
\label{se:apGreen}
The propagators needed to explicit the Wick contractions in the in-in formalism
must fulfill appropriate boundary conditions, corresponding to
\renewcommand{\arraystretch}{1.4}
\be
\label{eq:Greens}
\ba{rcl}
iG_A(t,\X)&=&\ds
\theta(-t)\pa{\Delta_+(t,\X)-\Delta_-(t,\X)}=
-i\frac{\delta(t+r)}{4\pi r}\,,\\
-iG_R(t,\X)&=&\ds 
\theta(t)\pa{\Delta_+(t,\X)-\Delta_-(t,\X)}=
i\frac{\delta(t-r)}{4\pi r}\,,\\
G_H(t-t',\X)&=&\ds\Delta_+(t,\X)+\Delta_-(t,\X)\,,
\ea
\ee
\renewcommand{\arraystretch}{1.}
the last equality in the first two lines of \ref{eq:Greens} holding for $d=3$ only.
$\Delta_\pm(t,\X)$ is given by
\be
\Delta_{\pm}(t,\X)=\int\frac{d^d\K}{(2\pi)^d}e^{\pm ikt}
\frac{e^{-i\K\cdot\X}}{2k}\,.
\ee

\subsection{The radiation reaction computation in coordinate space}
\label{ss:coord}
In order to compute the diagram in fig.~\ref{fig:radReac} the trilinear 
coupling of the gravitational field is needed. Using the metric ansatz in 
eq.~(\ref{eq:metric}), and the doubling of degrees of freedom required by the 
in-in formalism, the trilinear interactions from the Einstein-Hilbert Lagrangean
are: 
\renewcommand{\arraystretch}{1.4}
\be
\ba{rcl}
\mathcal{L}_{E-H} &\supset &\ds 
\left[-\frac{c_d}2 \dot\sigma_{+ij}\phi_+\dot\sigma_{-kl}
+ c_d\pa{\sigma_{+ij}\partial_k\phi_+\dot A_{-l}+
\sigma_{-ij}\partial_l\phi_+\dot A_{+k}} +\right.\\
&&\ds\left. \frac{c_d}2\pa{2\sigma_{+ij}\partial_k\phi_+\partial_l\phi_-+
\sigma_{-ij}\partial_k\phi_+\partial_l\phi_+}+
2c_d\, \partial_iA_{+j}\phi_+\partial_kA_{-l}\right]
\pa{\delta_{ik}\delta_{jl}+\delta_{il}\delta_{jk}-\delta_{ij}\delta_{kl}}\\
&&\left.\ds -2c_d\pa{A_{+i}\partial_i\phi_+\partial_t\phi_-+
A_{-i}\partial_i\phi_+\partial_t\phi_+}
-2c_d^2\partial_t\phi_+\phi_+\partial_t\phi_-\right]\,.
\ea
\ee
\renewcommand{\arraystretch}{1.}
In order to compute the amplitude one must Wick contract a quantity whose 
schematic structure is
\be
\ds(Q_-R_+&\ds +Q_+R_-)(t)(\Psi_{B-}\Psi_{C+}+\Psi_{C+}\Psi_{B-})(t_i)
(Q_-R_++Q_+R_-)(t')\phi_+(t_i)\phi_-(t'')
\ee
which gives
\be
\ba{rl}
\ds \ \bigg\{Q_-(t)&\left[-iG_R(t-t_i)\frac 12G_H(t_i-t')+
\frac 12D_H(t-t_i)(-iG_A(t_i-t'))\right.\\
&\ds\left.+\frac 12G_H(t-t_i)(-iG_A(t_i-t'))-
iG_R(t-t_i)\frac 12G_H(t_i-t')
\right]Q_-(t')\\
\ds +Q_-(t)&\paq{2(-iG_R(t-t_i))(-iG_R(t_i-t'))}Q_+(t')\\
\ds +Q_+(t)&\paq{2(-iG_A(t-t_i))(-iG_A(t_i-t'))}Q_-(t')\bigg\}
(-iG_R(t_i-t''))\,,
\ea
\ee
where tensor indices have been suppressed and we have assumed the field $R$ to
have non-vanishing contraction with any of the fields $\Psi_{B,C}$.
In our case we have to use the Riemann tensor expanded to first order as
\be
R^{0}_{\ i0j}=\frac 12\ddot\sigma_{ij}-\frac 12\dot A_{i,j}-\frac 12 \dot A_{j,i}
-\phi_{,ij}-\frac{\delta_{ij}}{d-2}\ddot\phi +O(h^2)\,,
\ee
where $h$ denotes generically the field $\phi$, $A$ or $\sigma$.
After taking the necessary Wick contractions and keeping
only the terms linear in $Q_-$ one obtains (in coordinate space, in $d=3$):
\renewcommand{\arraystretch}{1.5}
\be
\label{eq:apWick}
\ba{rl}
iS^{(MQ^2)}_{eff}&\ds = \frac 12\pa{\frac i{2\Lambda}}^2\pa{\frac{-iM}\Lambda}
\pa{\frac{i4\pi}\Lambda}\pa{\frac{(-i)^3}{4(8\pi)^3}}\int dt\,dt'\,dt_i
\int dr\, r^2\frac{d\Omega}{4\pi}\left\{\right.Q_{-ij}(t)Q_{+kl}(t')\times\\
&\ds\left[8\pa{\delta_{ij}\delta_{kl}-\delta_{ik}\delta_{jl}}\dpa_t^2\dpa_{t_i}\delta(t-t_i-r)
\dpa_{t_i}\dpa_{t'}^2\delta(t_i-t'-r)\frac 1{r^3}\right.\\
&\ds+ 32\delta_{jl}\pa{\dpa_i\frac 1r}\frac 1r\paq{\dpa_t^2\delta(t-t_i-r)
\dpa_{t_i}\dpa_{t'}\dpa_k\pa{\frac{\delta(t_i-t'-r)}r}}\\
&\ds -8\pa{\dpa_i\frac 1r}\frac 1r \dpa_t^2\delta(t-t_i-r)
\dpa_j\pa{\dpa_k\dpa_l+\delta_{kl}\dpa_{t'}^2}\pa{\frac{\delta(t_i-t'-r)}r}\\
&\ds +16\frac 1r\dpa_t\dpa_i\dpa_a\pa{\frac{\delta(t-t_i-r)}r}
\dpa_{t'}\dpa_k\dpa_b\pa{\frac{\delta(t_i-t'-r)}r}
\pa{\delta_{aj}\delta_{bl}-\delta_{al}\delta_{bj}+\delta_{ab}\delta_{jl}}\\
&\ds +8\pa{\dpa_i\frac 1r}\dpa_t\dpa_j\pa{\frac{\delta(t-t_i-r)}r}
\dpa_{t_i}\pa{\dpa_k\dpa_l+\delta_{kl}\dpa_{t'}^2}\pa{\frac{\delta(t_i-t'-r)}r}\\
&\ds\left.\left.-\frac 4r\dpa_{t_i}\pa{\dpa_i\dpa_j+\delta_{ij}\dpa_{t}^2}\pa{\frac{\delta(t-t_i-r)}r}\dpa_{t_i}
\pa{\dpa_k\dpa_l+\delta_{kl}\dpa_{t'}^2}\pa{\frac{\delta(t_i-t'-r)}r}\right]\right\}\,.
\ea
\ee
\renewcommand{\arraystretch}{1.}
which give the following contributions from respectively the $\sigma^2\phi$, 
$A^2\phi$, and $\phi^3$ vertices, which are the only ones contributing to the
logarithmic divergence ($u\equiv t-t'-2r$):
\renewcommand{\arraystretch}{1.5}
\be
\label{eq:effSapp}
\ba{rl}
\ds iS^{(MQ^2)}_{eff}=&\ds -i\frac{G_N^2M}4\int_{-\infty}^\infty dt\int_{-\infty}^\infty 
dt'\,Q_{-ij}(t)Q_{+kl}(t')
\int_0^\infty dr\,\frac 1{r^2}\times\\
&\ds\bigg\{\ \ \ 8\paq{r\delta^{(6)}(u)\pa{\delta_{ik}\delta_{jl}-
\delta_{ij}\delta_{kl}}}\\
&\ds\quad -\frac{16}{3}\left[r\delta^{(6)}(u)\delta_{ik}\delta_{jl}+
\pa{\delta_{ik}\delta_{jl}-\frac13\delta_{ij}\delta_{kl}}\times\right.\\
&\ds\qquad\left.\pa{6\delta^{(5)}(u)+15\frac{\delta^{(4)}(u)}{r}+
  18\frac{\delta^{(3)}(u)}{r^2}+9\frac{\delta^{(2)}(u)}{r^3}}\right]\\
&\ds\quad+\frac 8{15}\bigg[r\delta^{(6)}(u)\pa{\delta_{ik}\delta_{jl}+
13\delta_{ij}\delta_{kl}}\\
&\ds\qquad+\left.\pa{\delta_{ik}\delta_{jl}-\frac13\delta_{ij}\delta_{kl}}
\pa{6\frac{\delta^{(5)}(u)}{r}+15\frac{\delta^{(4)}(u)}{r^2}
+18\frac{\delta^{(3)}(u)}{r^3}+9\frac{\delta^{(2)}(u)}{r^4}}\right]\bigg\}
\,.
\ea
\ee
In order to obtain the final result we have to
repeatedly perform integration by parts, using
\be
\int_{-\infty}^t dt' \frac{Q(t')}{(t-t')^a}=
\left.\frac 1{a-1}\frac{Q(t')}{(t-t')^{a-1}}\right|^{t'=t}_{t'=-\infty}-
\frac 1{a-1}\int_{-\infty}^t dt'\,\frac{dQ(t')}{dt'}\frac 1{(t-t')^{a-1}}
\ee
and we imposed
\be
\left.\frac 1{a-1}\frac{Q(t')}{(t-t')^{a-1}}\right|^{t'=t}_{t'=-\infty}=0
\ee
for any $a\neq 1$, so discarding all but the logarithmic divergence, to
finally obtain
\be
iS_{eff}^{(MQ^2)}=-iG_N^2M
\frac 25\pa{\delta_{ik}\delta_{jl}+\delta_{il}\delta_{jk}-
\frac23\delta_{ij}\delta_{kl}}\int_{-\infty}^\infty dt\, Q_{-ij}(t)
\int_{-\infty}^t dt'\,Q_{+kl}^{(6)}(t')\frac 1{(t-t')}\,,
\ee
showing that the divergence is a short distance one, coming from the integration
region in which the (time) distance between the two quadrupole insertions goes
to zero.

\subsection{The radiation reaction computation in momentum space}

An analog computation can be performed in the Fourier space leading to
\be
\label{eq:Kspace}
\ba{rcl}
iS_{eff}^{4PN}&=&\ds
-i\frac M{16\Lambda^4}\int \frac{dk_0}{2\pi}
\pa{Q_{-ij}(k_0)\,Q_{+kl}(-k_0)+Q_{+ij}(-k_0)\,Q_{-kl}(k_0)}\\
&&\ds\int \frac{d^3k}{(2\pi)^d}\frac{d^3q}{(2\pi)^d}\,
\frac 1{\K^2-(k_0-ia)^2}\frac 1{(\K+\Q)^2-(k_0-ia)^2}\frac 1{\Q^2}\\
&&\ds\left\{
-\frac 18k_0^6\pa{\delta_{ik}\delta_{jl}+\delta_{il}\delta_{jk}-
  \frac 2{d-2}\delta_{ij}\delta_{kl}}\right.\\
&&\ds +\frac 12k_0^4\pa{q_iq_k\delta_{jl}}\\
&&\ds +\frac 1{2c_d}k_0^2\pa{-\frac{k_0^2q_iq_j}{d-2}\delta_{kl}
-q_iq_jk_kk_l-2k_iq_jk_kq_l-3q_iq_jq_kk_l-q_iq_jq_kq_l}\\
&&\ds \frac 12k_0^2\paq{k_ik_jq_kq_l-q_ik_jk_kq_l+
\delta_{ik}\pa{k_jk_l\K^2+k_jq_l(\K\Q)+\K^2k_jq_l+\K\Q k_jk_l}}+\\
&&\ds \frac 1{2c_d}k_0^2\pa{2q_ik_jq_kk_l-k_ik_jq_kq_l+q_ik_jq_kq_l
-k_0^2\delta_{kl}\frac{q_iq_j}{d-2}}\\
&&\ds-\frac 1{2c_d}k_0^2\left[k_ik_jk_kk_l+k_ik_jq_kq_l+2k_ik_jk_kq_l+\right.\\
&&\ds\qquad\left.\left.\frac {k_0^2}{d-2}\pa{\delta_{ij}k_kk_l+\delta_{kl}k_ik_j
+\delta_{ij}q_kq_l+2\delta_{ij}k_kq_l+\frac{k_0^2}{d-2}\delta_{ij}\delta_{kl}}
\right]\right\}
\,.
\ea
\ee
\renewcommand{\arraystretch}{1}
where we used the representation of the retarded(advanced) propagator
\be
G_{R(A)}=\lim_{a\to 0^+}\int \frac{dk_0}{2\pi}\frac{d^dk}{(2\pi)^d}e^{i(k_0t-\K\X)}
\frac 1{k^2-(k_0\mp ia)^2}\,.
\ee 
Defining $\int_q\int_k\equiv \int\frac{d^dq}{(2\pi)^d}\frac{d^dk}{(2\pi)^d}$
and $D\equiv\paq{\pa{(\K+\Q)^2-(k_0-ia)^2}\pa{k^2-(k_0-ia)^2}q^2}^{-1}$, all 
the integrals in eq.~(\ref{eq:Kspace}) can be evaluated
\be
\label{eq:int}
\ba{lcl}
\ds\int_q\int_kD&=&
\ds 2\frac{\Gamma^2(2-d/2)}{(4\pi)^d}
\frac{(-1)^{d-2}}{(d-2)(d-3)}\paq{\pa{k_0-ia}^2}^{d-3}\\
\ds\int_q\int_kk_ik_jq_kq_lD&=&\ds
\frac{(-1)^{d-2}\paq{\pa{k_0-ia}^2}^{d-1}}{(4\pi)^d}\frac{\Gamma^2(1-d/2)}
{2d(d-1)}\paq{\frac{2d}{d-2}\delta_{ij}\delta_{kl}+\frac{d-2}{d+2}
\pa{\delta_{ik}\delta_{jl}+\delta_{il}\delta_{jk}}}\\
\ds\int_q\int_kk_ik_jk_kk_lD&=&\ds 
\frac{(-1)^{d-2}\paq{\pa{k_0-ia}^2}^{d-1}\Gamma^2(2-d/2)}
{2(d-3)(d-2)^2(d-1)(4\pi)^d}\pa{1+2\frac{d-4}d+\frac{(d-4)(d-6)}{d(d+2)}}\\
&&\ds\pa{\delta_{ij}\delta_{kl}+\delta_{ik}\delta_{jl}+\delta_{il}\delta_{jk}}\\
\ds\int_q\int_kk_ik_jk_kq_lD&=&\ds 
\frac{(-1)^{d-2}[(k_0-ia)^2]^{d-1}}{(4\pi)^d}\frac{\Gamma(2-d/2)\Gamma(1-d/2)}
{2d(d-1)(d-2)}\pa{1+\frac{d-4}{d+2}}\\
&&\ds
\pa{\delta_{ij}\delta_{kl}+\delta_{ik}\delta_{jl}+\delta_{il}\delta_{jk}}\\
\ds\int_q\int_kk_ik_jD&=&\ds
\frac{\delta_{ij}}{(4\pi)^d}\frac{(-1)^d\paq{\pa{k_0-ia}^2}^{d-2}}{(d-2)^2(d-3)}
\Gamma^2(2-d/2)\paq{1+\frac{d-4}{d}}\\
\ds\int_q\int_kq_iq_jD&=&\ds
\frac{\delta_{ij}}{d(4\pi)^d}\Gamma^2(1-d/2)(-1)^{d-2}\paq{(k_0-ia)^2}^{d-2}\\
\ds\int_q\int_kk_iq_jD&=&\ds
\frac{\delta_{ij}}{2d(4\pi)^d}(-1)^{d-3}\paq{(k_0-ia)^2}^{d-2}\Gamma^2(1-d/2)\\
\ds\int_q\int_kq_iq_jq_kq_lD&=&\ds
\frac{(-1)^{d-2}\paq{(k_0-ia)^2}^{d-1}}{(4\pi)^d}\Gamma^2(-d/2)\frac d{2(d+2)}
\pa{\delta_{ij}\delta_{kl}+\delta_{ik}\delta_{jl}+\delta_{il}\delta_{jk}}\\
\ds\int_q\int_kq_iq_jq_kk_lD&=&\ds
(-1)^{d-2}\frac{(k_0-ia)^{2(d-1)}}{2(4\pi)^d(d+2)}\Gamma(1-d/2)\Gamma(-d/2)
\pa{\delta_{ij}\delta_{kl}+\delta_{ik}\delta_{jl}+\delta_{il}\delta_{jk}}
\ea
\ee
\renewcommand{\arraystretch}{1}
to obtain finally the result in eq.~(\ref{eq:radReacRes}).


\begin{thebibliography}{99}

\bibitem{BurkeThorne}
  W.~L.~Burke and K.~S.~Thorne, in ``Relativity", edited by M.~Carmeli, 
  S.~I.~Fickler and L.~Witten,
  (Plenum, New York, 1970) pp.209-228;\\
  W.~L.~Burke, J. Math. Phys. {\bf 12}, 401 (1971);\\
  K.~S.~Thorne, Astrophys. J. {\bf 158}, 997 (1969).

\bibitem{:2012dr}
  The LIGO Scientific Collaboration and the Virgo Collaboration,
  arXiv:1209.6533 [gr-qc].

\bibitem{Hulse:1974eb}
  R.~A.~Hulse, J.~H.~Taylor,
  Astrophys.\ J.\  {\bf 195 } (1975)  L51-L53.

\bibitem{Taylor:1982zz}
  J.~H.~Taylor, J.~M.~Weisberg,
  Astrophys.\ J.\  {\bf 253 } (1982)  908-920.

\bibitem{Blanchet_living}
  L.~Blanchet,
  Living Rev. Relativity 5,  (2002),  
  URL: 
  http://www.livingreviews.org/lrr-2002-3.

\bibitem{Blanchet:1987wq}
  L.~Blanchet, T.~Damour,
  Phys.\ Rev.\  {\bf D37 } (1988)  1410.

\bibitem{Blanchet:1993ng}
  L.~Blanchet,
  Phys.\ Rev.\  {\bf D47 } (1993)  4392-4420.

\bibitem{Blanchet:1993ec}
  L.~Blanchet and G.~Schaefer,
  Class.\ Quant.\ Grav.\  {\bf 10} (1993) 2699.

\bibitem{Blanchet:2010zd}
  L.~Blanchet, S.~L.~Detweiler, A.~Le Tiec, B.~F.~Whiting,
  Phys.\ Rev.\  {\bf D81 } (2010)  084033
  [arXiv:1002.0726 [gr-qc]].

\bibitem{Blanchet:2010cx}
  L.~Blanchet, S.~Detweiler, A.~Le Tiec and B.~F.~Whiting,
  Fundam.\ Theor.\ Phys.\  {\bf 162} (2011) 415
  [arXiv:1007.2614 [gr-qc]].

\bibitem{Goldberger:2004jt}
  W.~D.~Goldberger and I.~Z.~Rothstein,
  Phys.\ Rev.\  D {\bf 73} (2006) 104029
  [arXiv:hep-th/0409156].

\bibitem{EFToldcons}
  R.~A.~Porto,
  Phys.\ Rev.\  {\bf D73 } (2006)  104031
  [gr-qc/0511061];
  M.~Levi,
  Phys.\ Rev.\  {\bf D82 } (2010)  064029
  [arXiv:0802.1508 [gr-qc]];
  J.~B.~Gilmore and A.~Ross,
  Phys.\ Rev.\  D {\bf 78}, 124021 (2008)
  [arXiv:0810.1328 [gr-qc]];
  Y.~-Z.~Chu,
  Phys.\ Rev.\  {\bf D79 } (2009)  044031
  [arXiv:0812.0012 [gr-qc]];
  D.~L.~Perrodin,
  Proceedings of the MG12 Meeting on General Relativity, Paris, France, 12 - 18
  July 2009,
  arXiv:1005.0634 [gr-qc];
  R.~A.~Porto,
  Class.\ Quant.\ Grav.\  {\bf 27}, 205001 (2010)
  [arXiv:1005.5730 [gr-qc]];
  M.~Levi,
  Phys.\ Rev.\  {\bf D82 } (2010)  104004
  [arXiv:1006.4139 [gr-qc]];
  M.~Levi,
  Phys.\ Rev.\ D {\bf 85} (2012) 064043
  [arXiv:1107.4322 [gr-qc]].

\bibitem{Foffa:2011ub}
  S.~Foffa, R.~Sturani,
  Phys.\ Rev.\  {\bf D84 } (2011)  044031
  [arXiv:1104.1122 [gr-qc]].

\bibitem{Goldberger:2009qd}
  W.~D.~Goldberger and A.~Ross,
  Phys.\ Rev.\  {\bf D81 } (2010)  124015
  [arXiv:0912.4254 [gr-qc]];
\bibitem{EFTolddis}
  A.~Ross,
  Phys.\ Rev.\ D {\bf 85} (2012) 125033
  [arXiv:1202.4750 [gr-qc]].

\bibitem{EFTnewcons}
  R.~A.~Porto, I.~Z.~Rothstein,
  Phys.\ Rev.\ Lett.\  {\bf 97 } (2006)  021101
  [gr-qc/0604099];
  R.~A.~Porto, I.~Z.~Rothstein,
  Phys.\ Rev.\  {\bf D78 } (2008)  044012
  [arXiv:0802.0720 [gr-qc]]
  [Erratum-ibid.\  D {\bf 81} (2010) 029904];
  R.~A.~Porto, I.~Z.~Rothstein,
  Phys.\ Rev.\  {\bf D78 } (2008)  044013
  [arXiv:0804.0260 [gr-qc]]
  [Erratum-ibid.\  D {\bf 81} (2010) 029905];
  R.~A.~Porto,
  Phys.\ Rev.\  {\bf D73 } (2006)  104031
  [arXiv: gr-qc/0511061];
  M.~Levi,
  Phys.\ Rev.\ D {\bf 85} (2012) 064043
  [arXiv:1107.4322 [gr-qc]].

\bibitem{EFTnewdis}
  R.~A.~Porto,
  Phys.\ Rev.\ D {\bf 77} (2008) 064026
  [arXiv:0710.5150 [hep-th]];
  R.~A.~Porto, A.~Ross, I.~Z.~Rothstein,
  JCAP {\bf 1103 } (2011)  009
  [arXiv:1007.1312 [gr-qc]].
  W.~D.~Goldberger and I.~Z.~Rothstein,
  Phys.\ Rev.\ D {\bf 73} (2006) 104030
  [hep-th/0511133].

\bibitem{Porto:2012as}
  R.~A.~Porto, A.~Ross and I.~Z.~Rothstein,
  JCAP {\bf 1209} (2012) 028
  [arXiv:1203.2962 [gr-qc]].

\bibitem{Goldberger:2012kf}
  W.~D.~Goldberger, A.~Ross and I.~Z.~Rothstein,
  arXiv:1211.6095 [hep-th].

\bibitem{Foffa:2012rn}
  S.~Foffa and R.~Sturani,
  arXiv:1206.7087 [gr-qc].

\bibitem{EFTextmass}
  C.~R.~Galley, B.~L.~Hu,
  Phys.\ Rev.\  {\bf D79 } (2009)  064002
  [arXiv:0801.0900 [gr-qc]];
  C.~R.~Galley,
  [arXiv:1012.4488 [gr-qc]];
  C.~R.~Galley,
  Class.\ Quant.\ Grav.\  {\bf 29} (2012) 015011
  [arXiv:1107.0766 [gr-qc]].

\bibitem{Galley:2009px}
  C.~R.~Galley, M.~Tiglio,
  Phys.\ Rev.\  {\bf D79 } (2009)  124027.
  [arXiv:0903.1122 [gr-qc]].

\bibitem{Galley:2012qs}
  C.~R.~Galley and A.~K.~Leibovich,
  Phys.\ Rev.\ D {\bf 86} (2012) 044029
  [arXiv:1205.3842 [gr-qc]].

\bibitem{Galley:2012hx}
  C.~R.~Galley,
  arXiv:1210.2745 [gr-qc].

\bibitem{Kol}
  B.~Kol, M.~Smolkin,
  Class.\ Quant.\ Grav.\  {\bf 25 } (2008)  145011
  [arXiv:0712.4116 [hep-th]];
  B.~Kol, M.~Levi, M.~Smolkin,
  Class.\ Quant.\ Grav.\  {\bf 28 } (2011)  145021.
  [arXiv:1011.6024 [gr-qc]].

\bibitem{Schwinger:1960qe}
  J.~S.~Schwinger,
  J.\ Math.\ Phys.\  {\bf 2 } (1961)  407-432.

\bibitem{deWitt}  
  B.~DeWitt, Effectice action for expectation values, in \emph{Quantum concepts
  in Space and Time}, edited by R.~Penrose and C.~J.~Isham, Clarendon Press,
  Oxford, Clarendon, 1986.

\bibitem{Keldysh:1964ud}
  L.~V.~Keldysh,
  Zh.\ Eksp.\ Teor.\ Fiz.\  {\bf 47 } (1964)  1515-1527.

\bibitem{355730}
  B.~R.~Iyer and C.~M.~Will,
  Phys.\ Rev.\ Lett.\ \ {\bf 70} (1993) 113.


\bibitem{Jaranowski:2012eb}
  P.~Jaranowski and G.~Schafer,
  Phys.\ Rev.\ D {\bf 86} (2012) 061503
  [arXiv:1207.5448 [gr-qc]].

\bibitem{LeTiec:2011ab}
  A.~Le Tiec, L.~Blanchet and B.~F.~Whiting,
  Phys.\ Rev.\ D {\bf 85} (2012) 064039
  [arXiv:1111.5378 [gr-qc]].

\end{thebibliography}
\end{document}